\documentclass[a4paper,12pt]{article}
\pdfoutput=1
\usepackage{graphicx,rotating,hyperref,slashed,amsmath,xcolor,amssymb,amsfonts,expdlist,cite,charter}
\makeatletter
\hypersetup{colorlinks,bookmarksopen,bookmarksnumbered,
linkcolor=blus,pdfstartview=FitH,urlcolor=rossos,citecolor=verde}
\allowdisplaybreaks

\def\lsim{\mathrel{\rlap{\lower3pt\hbox{\hskip0pt$\sim$}}
   \raise1pt\hbox{$<$}}}         %less than or approx. symbol
\def\gsim{\mathrel{\rlap{\lower4pt\hbox{\hskip1pt$\sim$}}
   \raise1pt\hbox{$>$}}}         %greater than or approx. symbol

\newcommand{\mio}[1]{}

\newcommand{\fig}[1]{~\ref{fig:#1}}

\definecolor{Gray}{gray}{0.95}

\newcommand{\F}{\digamma}
\newcommand{\Q}{{\cal Q}}

\newcommand{\fb}{\,{\rm fb}}

\newcommand{\sfrac}[2]{#1/#2}

\usepackage{multicol}
\usepackage{color}
\definecolor{rosso}{cmyk}{0,1,1,0.4}
\definecolor{rossos}{cmyk}{0,1,1,0.55}
\definecolor{rossoc}{cmyk}{0,1,1,0.2}
\definecolor{blu}{cmyk}{1,1,0,0.3}
\definecolor{blus}{cmyk}{1,1,0,0.6}
\definecolor{bluc}{cmyk}{1,1,0,0.1}
\definecolor{verde}{cmyk}{0.92,0,0.59,0.25}
\definecolor{verdec}{cmyk}{0.92,0,0.59,0.15}
\definecolor{verdes}{cmyk}{0.92,0,0.59,0.4}

\newcommand{\riga}[1]{\noalign{\hbox{\parbox{\textwidth}{#1}}}\nonumber}

\newcommand{\eq}[1]{~{\rm (\ref{eq:#1})}}

\newcommand{\GeV}{\,{\rm GeV}}
\newcommand{\TeV}{\,{\rm TeV}}

\newcommand{\Tr}{\,{\rm Tr}}

\def\circa#1{\,\raise.3ex\hbox{$#1$\kern-.75em\lower1ex\hbox{$\sim$}}\,}

\newcommand{\beq}{\begin{equation}}
\newcommand{\eeq}{\end{equation}}
\newcommand{\mb}[1]{\mbox{\boldmath $#1$}}

\newcommand{\bea}{\begin{eqnarray}}
\newcommand{\eea}{\end{eqnarray}}
\newcommand{\be}{\begin{equation}}
\newcommand{\ee}{\end{equation}}
\font\tenrsfs=rsfs10 at 12pt
\font\sevenrsfs=rsfs7 at 10 pt
\font\fiversfs=rsfs5
\newfam\rsfsfam
\textfont\rsfsfam=\tenrsfs
\scriptfont\rsfsfam=\sevenrsfs
\scriptscriptfont\rsfsfam=\fiversfs
\def\mathscr#1{{\fam\rsfsfam\relax#1}}
\newcommand{\Ggg}{\Gamma_{\gamma\gamma}}
\def\Lag{\mathscr{L}}

\def\circa#1{\,\raise.3ex\hbox{$#1$\kern-.75em\lower1ex\hbox{$\sim$}}\,}
\makeatletter

\def\hhref#1{\href{http://arxiv.org/abs/#1}{arXiv:#1}} % in bibliography

\newcommand{\doi}[1]{\href{http://dx.doi.org/#1}{[doi]}}

\def\hhref#1{\href{http://arxiv.org/abs/#1}{arXiv:#1}} 
 
\def\art{\@ifnextchar[{\eart}{\oart}}
\def\eart[#1]#2#3#4#5#6{{\rm #2}, {\em #3 \bf #4} {\rm (#6) #5} ({\em #1})}

\def\article{\@ifnextchar[{\earticle}{\oarticle}}
\def\oarticle#1#2#3#4#5#6{{\rm #1}, {\em ``#6''}, {\rm #2 #3 (#5) #4}}
\def\earticle[#1]#2#3#4#5#6#7{{\rm #2}, {\em ``#7''}, {\rm #3 #4 (#6) #5}  [\hhref{#1}]}
\def\hepart[#1]#2{{\rm #2, \em#1}}
\def\heparticle[#1]#2#3{#2, {\em ``#3''} [\hhref{#1}]}

\newcounter{alphaequation}[equation]
\def\thealphaequation{\theequation\hbox to
0.6em{\hfil\alph{alphaequation}\hfil}}
\def\eqnsystem#1{
\def\@eqnnum{{\rm (\thealphaequation)}}
\def\@@eqncr{\let\@tempa\relax \ifcase\@eqcnt \def\@tempa{& & &} \or
  \def\@tempa{& &}\or \def\@tempa{&}\fi\@tempa
  \if@eqnsw\@eqnnum\refstepcounter{alphaequation}\fi
\global\@eqnswtrue\global\@eqcnt=0\cr}
\refstepcounter{equation} \let\@currentlabel\theequation \def\@tempb{#1}
\ifx\@tempb\empty\else\label{#1}\fi
\refstepcounter{alphaequation}
\let\@currentlabel\thealphaequation
\global\@eqnswtrue\global\@eqcnt=0 \tabskip\@centering\let\\=\@eqncr
$$\halign to \displaywidth\bgroup \@eqnsel\hskip\@centering
$\displaystyle\tabskip\z@{##}$&\global\@eqcnt\@ne
\hskip2\arraycolsep\hfil${##}$\hfil& \global\@eqcnt\tw@\hskip2\arraycolsep
$\displaystyle\tabskip\z@{##}$\hfil
\tabskip\@centering&\llap{##}\tabskip\z@\cr}
\def\endeqnsystem{\@@eqncr\egroup$$\global\@ignoretrue} \makeatother

\newcommand{\SU}{\,{\rm SU}}

\newcommand{\U}{\,{\rm U}}

\definecolor{fiorentina}{rgb}{.5,0,.5}

\begin{document}

\vspace{-0.2truecm}

\centerline{CERN-PH-TH-2016-128}

\vspace{0.4truecm}

\begin{center}
\boldmath

{\textbf{\LARGE\color{magenta} The Higgs of the Higgs and the diphoton channel}}
\unboldmath

\bigskip

%\vspace{0.4truecm}
\vspace{0.4truecm}

{\bf Kristjan Kannike,$^{a}$ Giulio Maria Pelaggi,$^{b}$ Alberto Salvio,$^{c}$ Alessandro Strumia$^{a,b,c}$}
 \\[5mm]

{\it $^{a}$ National Institute of Chemical Physics and Biophysics, R\"avala 10, 10143 Tallinn, 
Estonia}
\\[2mm]
{\it $^b$ Dipartimento di Fisica dell'Universit{\`a} di Pisa and INFN, Italy}
\\[2mm]
{\it $^c$ CERN, Theory Division, Geneva, Switzerland}\\[1mm]

\vspace{2cm}

\thispagestyle{empty}
{\large\bf\color{blus} Abstract}
\begin{quote}\large
LHC results do not confirm
conventional natural solutions to the Higgs mass hierarchy problem,
motivating alternative interpretations
where a hierarchically small weak scale is generated from a dimension-less quantum dynamics.
We propose weakly and strongly-coupled models where
the field that breaks  classical scale invariance giving mass to itself and to the Higgs
is identified with a possible new resonance within the LHC reach. As an example, we identify such resonance with the 750 GeV diphoton excess recently reported by ATLAS and CMS.
Such models can be extrapolated up to the Planck scale, provide
Dark Matter candidates and eliminate the SM vacuum instability.
%and can provide a first order phase transition.
\end{quote}
\thispagestyle{empty}
\end{center}

\setcounter{page}{1}
\setcounter{footnote}{0}

\newpage

\tableofcontents

%\newpage

\section{Introduction}
Conventional solutions to the
hierarchy problem have been disfavoured by experimental bounds that relegated them to fine-tuned corners of their parameter space.
This situation motivated a reconsideration of the hierarchy problem:
no physical effects are associated with quadratically divergent corrections to the Higgs squared masses,
so maybe we are over-interpreting quantum field theory when we require theories that tame such quadratic divergences.

The hierarchy problem is bypassed if there are no particles much heavier than the Higgs and significantly coupled to it,
such that physical corrections to the Higgs mass are naturally small. This heretic context was dubbed `finite naturalness' in~\cite{1303.7244}.

Within the general finite naturalness scenario  an interesting sub-set of theories are those described by dimension-less Lagrangians, 
that assume  that massive parameters do not exist at fundamental level,
such that all mass scales in nature are generated dynamically, like the QCD scale %$\Lambda_{\rm QCD}$ 
in the Standard Model.
Strictly speaking, finite-naturalness does not necessarily require the absence of masses in the Lagrangian. What makes the dynamical generation of masses attractive is the fact that it can lead to a separation of scales, which depends exponentially on dimensionless couplings.
% (just like in QCD when $\Lambda_{\rm QCD}$ is dynamically generated).
Therefore, this specific setup allows us to justify why the weak scale is many orders of magnitude smaller than the Planck scale,
which itself might be generated by dimension-less dynamics,
with important implications for inflation~\cite{agravity}.

Notice that a dimension-less Lagrangian  corresponds to a classical scale invariant model. We do not wish to preserve scale invariance at the quantum level because we eventually have to generate the observed scales. When the couplings of the theory are small, classical scale invariance is also an approximate symmetry of the model, otherwise it is just the requirement that the masses are dynamically generated (with the motivations stated above). 

Various models where the weak scale  arises in this way have been proposed.
They can be classified in two categories:
\begin{enumerate}

\item The weak scale is the scale where a scalar quartic  coupling $\lambda$ runs negative, inducing vacuum expectation values
{\em \`a la} Coleman-Weinberg~\cite{1306.2329,lambdamod,gmod}.

\item The weak scale is the scale where a gauge coupling $g$ runs to non-perturbative values, inducing condensates~\cite{1410.1817,strong}.

\end{enumerate}

It is interesting to ask whether these theoretical frameworks could lead to something visible at the LHC. A particularly clean channel, which has therefore a great discovery potential, is the one with two photons  in the final state. We therefore study models where one of the required new particles can be identified with a new resonance $\digamma$ decaying into two   photons (diphoton). As an example, we treat in some detail the case in which such resonance is identified with the recent diphoton excess at 750 GeV reported by ATLAS and CMS \cite{data}. Other numerical choices are of course possible and our work will remain valuable even if such excess will turn out to be a statistical fluctuation.
 
 %The 750 GeV digamma resonance  ($\digamma$), if real, raises the question: ``who ordered that?''
%Presumably it has something to do with the origin of the electro-weak scale.
Connections with conventional
 solutions to the hierarchy problem have been explored by authors who tried
to identify $\F$ with one or another supersymmetric particle~\cite{1512.04933,susy}
or with resonances of composite Higgs scenarios~\cite{1512.04933,CH}.

References~\cite{1512.06708,1512.09136} tried to incorporate the diphoton hinted by ATLAS and CMS in the weakly-coupled framework;
however, in both cases couplings are so large that
the attempted models hit Landau poles just above the weak scale, 
such that no hierarchy is dynamically generated.\footnote{
Furthermore~\cite{1512.06708} also contain some explicit mass term.
The model in~\cite{1512.07225} can be extrapolated up to infinite energy, but it employs explicit mass terms.  It is possible that these mass terms could be generated at a scale much higher than the weak scale. But finding an example of this sort goes beyond the scope of the present paper.}

\medskip

In section~\ref{weak} we present weakly-coupled dimensionless models
where $\digamma$ is the field that dynamically generates the weak scale, while running down from the Planck scale.
%that reproduce the diphoton excess as a narrow resonance
%and that can be extrapolated up to the Planck scale.
%This can be easily done if the diphoton is an extra particle added ad-hoc.
%More interestingly, in section~\ref{weak} we show that diphoton can be 
%as the particle that dynamically generates the weak scale.
In section~\ref{strong} we present strongly coupled dimension-less models,
where, among other things, the diphoton excess is reproduced.
In section~\ref{concl} we present our conclusions.

%Previous attempts failed: Sannino~\cite{1512.06708}
%[The problem of the model is that their CW needs
%a large quartic $\approx 5$ such the model becomes non-perturbative at a TeV: no
%mass hierarchy is dynamically generated.
%Furthermore they have explicit masses for the fermion mediators.
%]
%and~\cite{1512.09136}
%[But they have no connection with the weak scale and they allow large $\F$ vev, up to $10^7\GeV$,
%such that it can be light only by being very weakly coupled: 
%$\F\to\gamma\gamma$ is negligible.
%They avoid this adding a massive cubic, so it's not dimensionless.]

%\vspace{1cm}
%\xxx{Alberto: aggiungere una review su CW?}

\section{Weakly coupled models}\label{weak}
Various extensions of the SM where the weak scale is generated {\em \`a la} Coleman-Weinberg have
been proposed in the literature. They can be divided into two main categories,
depending on which correction renormalises a quartic coupling $\lambda_\F$
down to negative values at low energy, such that the dimension-less potential $\lambda_\F \F^4$
develops a minimum: either
%\beq 
%\frac{d}{dt}\lambda = \lambda (s_\lambda  \lambda + s_{\lambda y}  y^2 - s_{\lambda g}  g^2) -s_y y^4 + s_g  g^4 \ ,
%\label{eqlam0}
%\eeq
\begin{itemize}
\item[$\lambda$)] corrections due to other scalar quartics~\cite{lambdamod}; or
\item[$g$)]  corrections due to a gauge coupling~\cite{1306.2329,gmod}.\footnote{Yukawa couplings have the opposite effect of making a quartic larger at low energy.}
%(with extra Yukawas $\lambda$ can run touching zero generating a minimum with zero cosmological constant).
\end{itemize}
Models of type $\lambda)$ are more problematic than models of type $g)$, 
because a non-abelian gauge coupling $g$ can be sizeable without implying nearby Landau poles,
while a sizeable quartic coupling drives itself to larger values at higher energies.
Models of type $\lambda)$ in the literature mitigate this effect by having a large number $N$ of smaller quartics,
and we will follow this strategy.
We present in section~\ref{FN} (section~\ref{FHH})
models of type $g)$ (of type $\lambda$) where the
Coleman-Weinberg field is the diphoton, and where the RGE can be extrapolated up to the Planck scale,
such that a large hierarchy is dynamically generated.

\bigskip

Before starting, we mention a broader --- but less interesting --- class of scale-invariant models,
%\subsection{Models where the diphoton is an extra ad-hoc scalar}\label{adhoc}
where the diphoton is added as an ad-hoc extra field that does not play a key role in the dynamical generation of the weak scale.
Roughly, one can choose any one of the $N$ diphoton models proposed in the literature,
and any one of the $M$ models that dynamically generated the weak scale,
and combine them into $N\times M$ models.

For example, there is no obstacle in combining the dimension-less model in~\cite{1306.2329}
(an extra $\SU(2)$ gauge interaction
with an extra scalar doublet $S$, that acquires a vacuum expectation value
through the Coleman-Weinberg mechanism)
with the `diphoton everybody's model'~\cite{1512.04933}
(the diphoton is an extra scalar singlet $\F$ coupled to extra charged scalars $X$ or fermions $\psi$).
%The diphoton mass and the $X$ mass can arise from quartic couplings to $S$;
%alternatively the fermion mass can arise from a Yukawa coupling to $\F$, provided that its squared mass is negative such that
%it acquires a vacuum expectation value.
The SU(2) massive vectors are DM candidates~\cite{1306.2329}.

What are the generic features of this class of ad-hoc models?
The new charged particles cannot be arbitrarily heavier
than the Higgs mass $M_h$, otherwise they contribute to it unnaturally~\cite{1303.7244}:
\beq \begin{array}{ll}
M_\psi \circa{<} 4\pi M_h/g_{1,2}^2\qquad &\hbox{for fermions,}\\
M_X\circa{<} M_h/g^2_{1,2}&\hbox{for scalars,}
\end{array}\eeq
where $g_{1,2}$ are the electroweak gauge couplings.
%(given that RGE effects induce
%$\lambda_{HS}  \circa{>} \lambda_{HX}\lambda_{SX}$ and  $\lambda_{HX}\circa{>} g^4_{Y,2}$,
%having assumed that RGE logarithms $\ln (M_{\rm Pl}/M_h)$ are large enough to compensate for the loop factor $\sim 1/(4\pi)^2$).
The measured diphoton rate indeed suggests new particles with sub-TeV mass~\cite{1512.04933}.
Furthermore, in the context of finite naturalness,
the existence of extra colored fermions below a few TeV is necessary
if the QCD $\theta$ problem is solved by an KSVZ axion model~\cite{1303.7244}.

\medskip

We now propose more interesting --- but  more constrained --- models where
the diphoton $\F$ is the particle that dynamically acquires the vacuum expectation value that
 induces the electroweak scale.
%This identification is not immediate, because such particle needs to couple to the Higgs boson
%(thereby mixing with it, acquiring a decay width into $WW,ZZ$), while the diphoton excess needs a coupling to $\gamma\gamma$.

\begin{table}
$$\begin{array}{|cc|cccc|}\hline
\hbox{name} &\hbox{spin}   & \SU(3)_c & \SU(2)_L & \U(1)_Y & \SU(N)  \\
\hline 
%H &0& 1& 2 & -1/2 & N   \\
X & 0 &1 &1 & 0 &  1\\
S & 0 & 1 &1 & 0 &  N\\
\hline
{\cal U} & 1/2 & \bar 3 &1 &-2/3 & N\\
{\cal U}^c & 1/2 &  3 &1 &+2/3 &\bar N\\ \hline
\end{array}$$
\caption{\label{tab:M1}\em Beyond the Standard Model
field content of the model of type $g)$ of section~\ref{FN}.}
\end{table}

\subsection{Model of type $g$)}\label{FN}
%We consider a model where
%the scalar diphoton is neutral under the SM gauge group, 
%but charged under the 
We extend the SM by adding an extra gauge group $\SU(N)$ and the extra fields
listed in table~\ref{tab:M1}:
two scalars $S$ and $X$, a quark ${\cal U}$ with the same quantum numbers of the SM right-handed up quarks
that fill a $N$ of $\SU(N)$, plus the conjugated fermion ${\cal U}^c$ in order to form a vector-like quark.
The  dimension-less Yukawa couplings are
\beq \Lag_Y = \Lag_Y^{\rm SM}+ (y_S S {\cal U}^c U + y_X\, X  {\cal U}{\cal U}^c +\hbox{h.c.}),\eeq
where $U$ is the right-handed up quark of the SM.
The dimension-less potential of the theory is
\beq
 V(H,S,X) = \lambda_H |H|^4 
+\lambda_X X^4+ \lambda_S |S|^4
 - \lambda_{HS} |H|^2 |S|^2
 - \lambda_{HX} |H|^2 X^2 -\lambda_{SX} |S|^2 X^2  . \label{eq:V2}\eeq
The tree-level potential is positive, $V\ge 0$, when the quartic couplings satisfy~\cite{Kannike:2012pe}
\beq\left\{
\begin{array}{l}
  \lambda_{H} \ge 0, \qquad  \lambda_{S} \ge 0, \qquad \lambda_{X} \ge 0,
  \\
  \bar{\lambda}_{HS} \equiv -\lambda_{HS} + 2 \sqrt{\lambda_{H} \lambda_{S}} \ge 0,
  \\
  \bar{\lambda}_{HX} \equiv -\lambda_{HX} + 2 \sqrt{\lambda_{H} \lambda_{X}} \ge 0,
  \\
  \bar{\lambda}_{SX} \equiv -\lambda_{SX} + 2 \sqrt{\lambda_{S} \lambda_{X}} \ge 0,
  \\
 -\sqrt{\lambda_{H}} \lambda_{XS} - \sqrt{\lambda_{S}} \lambda_{HX} - \sqrt{\lambda_{X}} \lambda_{HS} 
  + 2 \sqrt{\lambda_{H} \lambda_{S} \lambda_{X}} + \sqrt{\bar{\lambda}_{HS} \bar{\lambda}_{HX} \bar{\lambda}_{SX}} \ge 0.
\end{array}\right.\label{eq:V2p}\eeq
The RGEs of the model are listed in appendix~\ref{RGEg}.
Notice that $\lambda_{HS}$ is unavoidably generated from the Yukawa couplings,
if  the top is the quark mostly coupled to the new states ${\cal U}$.

\subsubsection*{Masses}
\begin{figure}[t]
$$\includegraphics[width=0.76\textwidth]{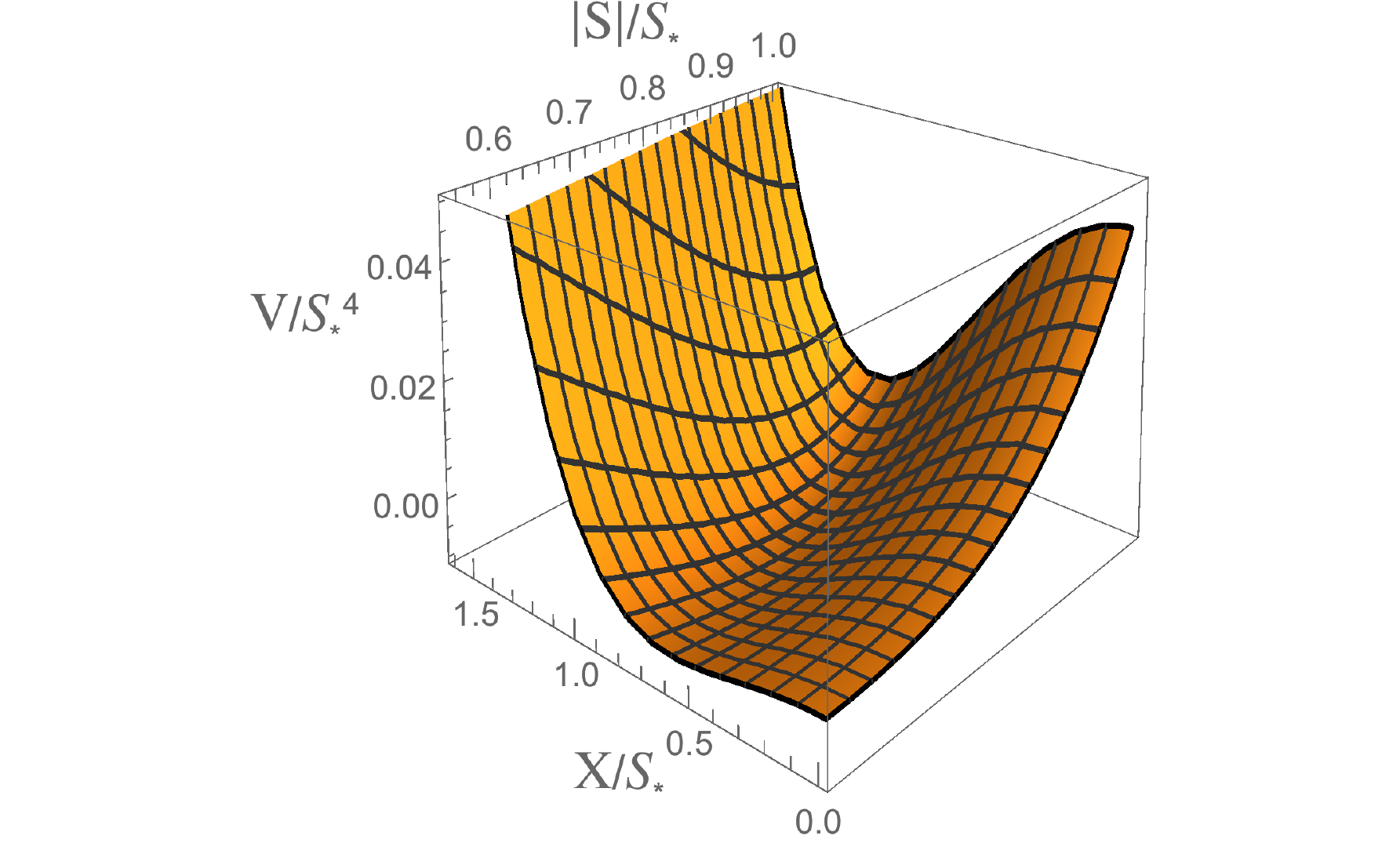}$$ 
\caption{\em 
\label{fig:Run2} Potential defined in eq.~(\ref{eq:V2}) for $\lambda_S$ chosen as in eq.~(\ref{lambdaSS}) and  $H=0$. We set  $\beta_{\lambda_S}\simeq 0.12$, $\lambda_X\simeq 0.030$ and $\lambda_{SX}\simeq 0.071$. 
\label{PlotPot}}
\end{figure}

When the field $S$ dynamically acquires a vacuum expectation value breaking
$\SU(N)\to\SU(N-1)$ (to nothing if $N=2$~\cite{1306.2329}),
$H$ and $X$ too can acquire vacuum expectation values, in view of their $\lambda_{HS}$ and $\lambda_{SX}$ quartic couplings to $S$.
In the unitary gauge there are three physical scalars:
\beq S = ( v_S+\frac{s}{\sqrt{2}},0,\ldots),\qquad X = v_X + x,\qquad H = (v+\frac{h}{\sqrt{2}},0).\eeq
We assume that $\lambda_{HS}$ and $\lambda_{HX}$ are negligible, and that
scale invariance gets broken when the RGE running of $\lambda_S$, dominated by the gauge coupling $g$, 
violates the condition $\bar\lambda_{SX}\ge 0$ of eq.\eq{V2}.
We can approximate the one loop potential by inserting a running
\beq \label{lambdaSS} \lambda_S(S) =\frac{ \lambda_{SX}^2}{4\lambda_X} + \beta_{\lambda_S} \ln\frac{|S|}{S_*}\eeq
in the tree-level potential of eq. \eq{V2}. Here $S_*$ is the scale at which the stability condition $\bar{\lambda}_{SX}\geq 0$ is violated. For $|S|$ around $S_*$ there is an approximately flat direction, which is lifted by quantum correction (the log term in eq. (\ref{lambdaSS})) as illustrated in fig.\ref{PlotPot}.
This generates the following absolute minimum of the potential 
 \beq 
{v_S} =\frac{S_*}{e^{1/4}},\qquad
v_X = v_S \sqrt{\frac{\lambda_{SX}}{2\lambda_X} },
 \eeq
 which is visible in fig.\ref{PlotPot}.
 There is another absolute minimum with the sign of $v_X$ switched,  but we assume $v_X > 0$ without loss of generality.
The scalar mass matrix at the minimum in the ($x,s$) basis is
\beq 2v_S^2
\begin{pmatrix}
2\lambda_{SX} & - \sqrt{\lambda_{SX}^3/\lambda_X} \cr
- \sqrt{\lambda_{SX}^3/\lambda_X}  & \beta_{\lambda_{S}}+\lambda_{SX}^2/2\lambda_X
\end{pmatrix}.
\eeq
Neglecting the small mixing with the Higgs,
the mass eigenstates are the diphoton $\F$ and a similar heavier scalar $\F'$
\beq\left\{\begin{array}{l}
\F = s \cos\theta+ x \sin\theta \cr
\F' = x\cos\theta-s\sin\theta
\end{array}\right.\qquad\hbox{where}
\qquad \tan2\theta = \frac{4\lambda_{SX}^{3/2}\lambda_X^{1/2}}{4\lambda_X \lambda_{SX}-
\lambda_{SX}^2-2 \lambda_X \beta_{\lambda_S}}.\eeq
Their masses are
\beq M_{\F,\F'}^2 = \frac{v_S^2}{2\lambda_X}\bigg[A \mp \sqrt{A^2 - 32\beta_{\lambda_S}\lambda_X^2\lambda_{SX}}\bigg],\qquad
A\equiv 4\lambda_X \lambda_{SX}+
\lambda_{SX}^2+2 \lambda_X \beta_{\lambda_S}.\eeq
The Higgs mass is $M_h^2 \simeq 2(\lambda_{HS} v_S^2 + \lambda_{HX} v_X^2)$.
The mixing angle between the Higgs and the diphoton  is experimentally constrained to be small~\cite{1512.04933,1604.06446}
\beq |\sin\theta_{h\F}| \circa{<} 0.015 \sqrt{\sfrac{\Ggg}{10^{-6}M_\F}}. \label{eq:hFmixexp}\eeq
In the present model such mixing angle is of order $ \theta_{h\F} \sim \sfrac{M_h^2 v}{M_\F^2 v_{S,X}}$,
which is below its experimental bound provided that $v_{S,X}\circa{>}500\GeV$.

\medskip

Coming to fermion masses,
 ${\cal U}$ and ${\cal U}^c$  split into $N-1$ vector-like up quarks with mass
$M_{N-1}=y_X v_X $ and into one with mass
$M_1=\sqrt{(y_S v_S)^2 + (y_X v_X)^2}$, having neglected smaller electroweak contributions.
Recasting LHC searches for similar objects~\cite{ATLAS}, we estimate a bound $M\circa{>}1.2 ~(1.3)\TeV$ on their masses for $N=1~(5)$.

\subsubsection*{Diphoton rate}
The ${\cal U}$ heavy quarks mediate diphoton decays into SM vectors.
%At one loop, such fermions induce $\F$ decays into SM vectors.
%The SM vectors acquire couplings of the form
%$SS^\dagger F_{\mu\nu}^2$ and $XF_{\mu\nu}^2$.
%$S$ and $X$ contain two scalars, that mix among themselves.
In the limit where the particles $\wp$ that induce loop decays of a generic scalar $\varphi$
into SM vectors are much heavier than 
the energy involved in the decay, their contribution to the decay amplitude is  related to their
contributions to the $\beta$ function coefficients $\Delta b_i^\wp$ of the SM gauge couplings
 as~\cite{LET}
\beq\label{eq:LET}
 \Lag_{\rm eff} = \sum_{i,\wp} 
   \Delta b_{i}^\wp \frac{\alpha_i}{8\pi} (F_{\mu\nu}^{i})^2 \ln\frac{M_\wp(\varphi)}{M_\wp}
%= c_{\gamma\gamma}e^2 \frac{\F F_{\mu\nu}^2}{2M_\F} + \cdots\qquad
%c_{\gamma\gamma} =\sum_X  \frac{\Delta b^X_{\rm em} }{8\pi^2} \frac{M_\F}{w} 
\eeq
where 
$M_\wp(\varphi) $ is the $\wp$ mass for a generic vev of $\varphi$.
In our model $\varphi=\{x,s\}=\{\F,\F'\}$ and the loop particles $\wp$ are the quark triplets
with mass $M_{N-1}=y_X X$ ($\Delta b_{\rm em} = 16/9$) and with mass $M_1=\sqrt{y_X X^2 + y_S |S|^2}$ ($\Delta b_{\rm em} = (N-1)16/9$).
The other $\Delta b_i$ coefficients are given by eqs.\eq{db1} and\eq{db3}. 
So
\beq \Lag_{\rm eff}^\gamma =\frac{\alpha_{\rm em}}{9\pi} F_{\mu\nu}^2\bigg[\sqrt{2}s y_S\frac{\sqrt{M_1^2 - M_{N-1}^2} }{M_{1}^2}+2
x y_X\frac{(N-1) M_1^2+M_{N-1}^2}{M_1^2 M_{N-1}}\bigg].\eeq
Rotating to the mass eigenstates one finds the $\F$ width into $\gamma\gamma$:
\begin{eqnsystem}{sys:Ggg2}
\label{eq:GggLET2}
\frac{\Gamma(\F\to\gamma\gamma)}{M_\F} &=& \frac{4 M_\F^2 \alpha_{\rm em}^2}{81\pi^3}
\bigg(   \frac{\sqrt{2}y_S^2  v_Sv_X \cos\theta +2 (N y_X^2 v_X^2 +(N-1) y_S^2 v_S^2) \sin\theta}{v_X(y_S^2 v_S^2  + y_X^2 v_X^2) }  \bigg)^2,\\
\label{eq:GggLET2bis}
\frac{\Gamma(\F'\to\gamma\gamma)}{M_{\F'}} &=& \frac{4 M_{\F'}^2 \alpha_{\rm em}^2}{81\pi^3}
\bigg(   \frac{\sqrt{2}y_S^2  v_Sv_X \sin\theta -2 (N y_X^2 v_X^2 +(N-1) y_S^2 v_S^2) \cos\theta}{v_X(y_S^2 v_S^2  + y_X^2 v_X^2) }  \bigg)^2.
\end{eqnsystem}
In the limit of small $\beta_{\lambda_S}$ the pseudo-Goldstone of
scale invariance is the lighter state $\F$:
\beq \label{eq:massless}
M_\F^2\simeq \frac{2v_S^2  \beta_{\lambda_S}}{1+\sfrac{\lambda_{SX}}{4\lambda_X}} \ll 
M_{\F'}^2 \simeq 4 v_S^2\lambda_{SX} \left( 1+\frac{\lambda_{SX}}{4\lambda_X} \right),
\qquad\tan\theta = \frac{v_X}{{\sqrt{2}}v_S}\simeq \sqrt{\frac{\lambda_{SX}}{2\lambda_X}}\eeq
and eq.\eq{GggLET2} reduces to
\beq\label{eq:GggLET2lim}
\frac{\Gamma(\F\to\gamma\gamma)}{M_{\F}} \simeq \frac{8N^2 M_\F^2 \alpha_{\rm em}^2}{81\pi^3 (v_S^2 + v_X^2/2)}.\eeq
In the less relevant opposite limit of small $\lambda_{SX}$ (and thereby $v_X\ll v_S$)
the pseudo-Goldstone of
scale invariance is the heavier scalar $\F'$:
\beq M_\F^2\simeq 4 v_S^2\lambda_{SX}\ll
M_{\F'}^2 \simeq 2v_S^2  \beta_{\lambda_S} ,
\qquad\tan\theta \simeq \beta_{\lambda_S} \sqrt{\frac{\lambda_{X}}{\lambda^3_{SX}}}\eeq
and eq.\eq{GggLET2bis} reduces to
\beq\frac{\Gamma(\F'\to\gamma\gamma)}{M_{\F'}} \simeq \frac{8 M_{\F'}^2 \alpha_{\rm em}^2}{81\pi^3 v_S^2 }.\eeq
Going beyond the approximation of eq.\eq{LET} requires computing Feynman diagrams
with two different masses in the loop.
In the limit $y_S=0$, such that $M_1=M_{N-1}$, the full expression is
\beq \frac{\Gamma(\F\to\gamma\gamma)}{M_\F}=
\frac{4N^2 M_{1}^2 \alpha^2_{\rm em} y_X^2 \sin^2\theta }{9\pi^3M_\F^2} \left|{\cal S}\left(\frac{4 M_{1}^2}{M_\F^2}\right)\right|^2 
\label{eq:gamrat2}\eeq
where the loop function ${\cal S}$ is
\beq {\cal S}(x) = 1 + (1-x) \arctan^2\left(\frac{1}{\sqrt{x-1}}\right)\stackrel{x\gg1}{\simeq} \frac{2}{3x}.\eeq
In the limit $M_1\gg M_\F$ eq.\eq{gamrat2} reduces to\eq{GggLET2lim}.

\begin{figure}[t]
$$\includegraphics[width=0.46\textwidth]{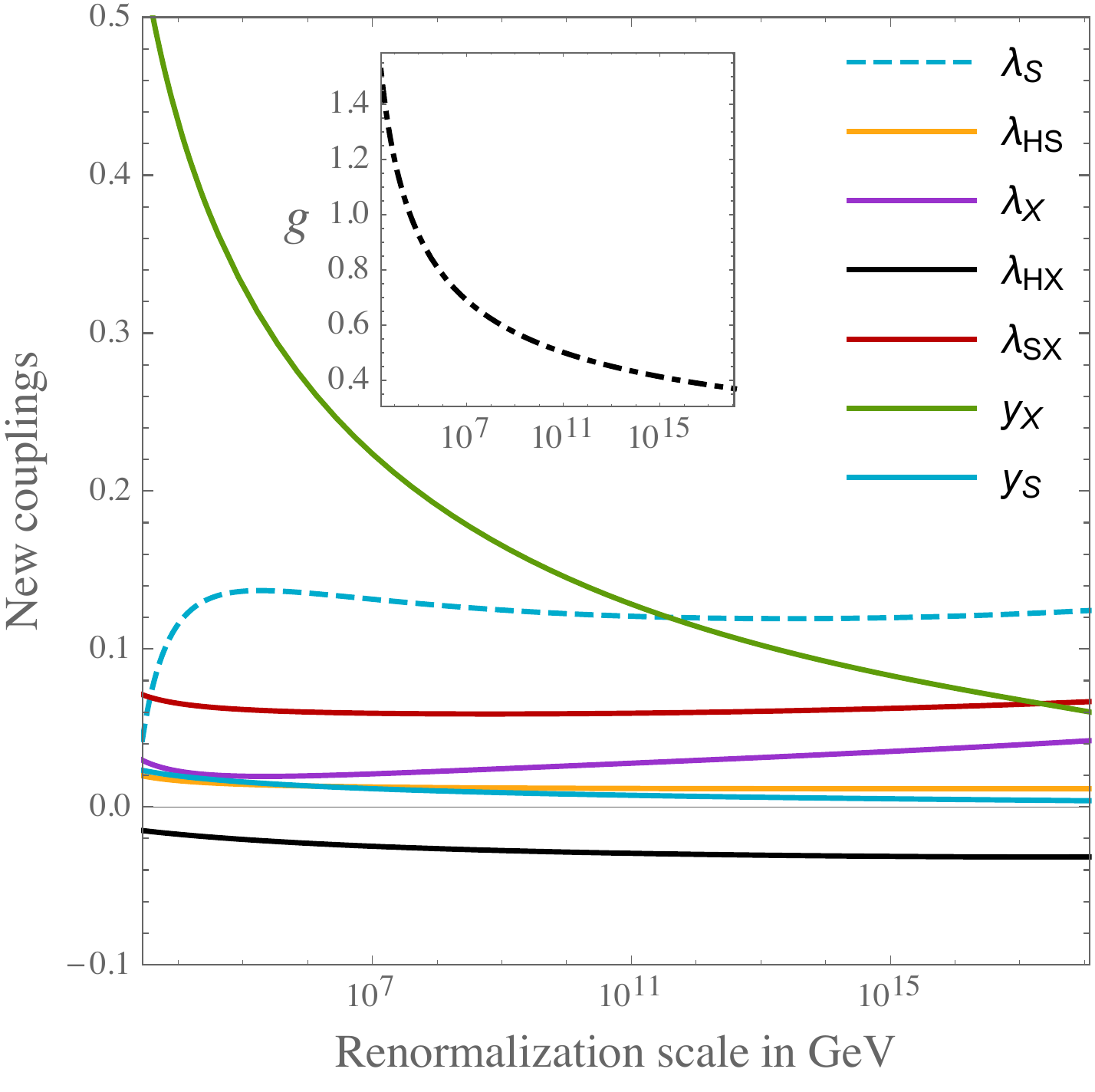}\qquad\qquad
\includegraphics[width=0.44\textwidth]{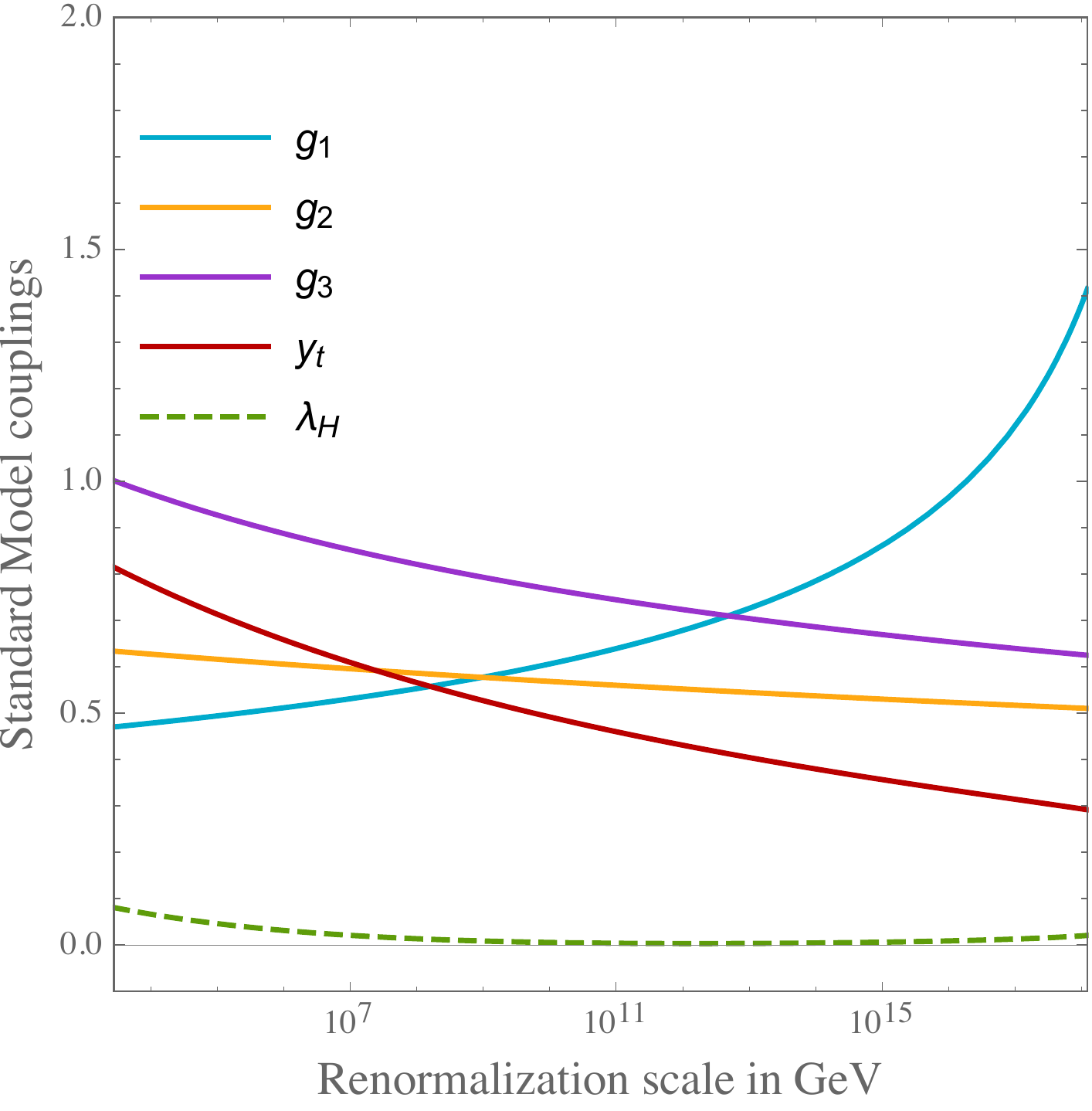}$$ 
\caption{\em 
\label{fig:Run2} Running of the couplings for  central values for the SM parameters and for  $N=5$, $g(S_*) \simeq 1.5$, $v_S\simeq 2.2 \TeV$,  $\lambda_{HX}(S_*) \simeq -0.015$, $\lambda_{SX}(S_*) \simeq 0.071$, $\lambda_{X}(S_*) \simeq 0.030$, $y_{S}(S_*)= 0.023$ and  $y_{X}(S_*)= 0.53$. 
\label{run2}}
\end{figure}

\subsubsection*{Numerical example}
By performing a global fit of  run 1 and run 2 ATLAS and CMS data 
assuming a narrow width and dominant $gg\to\F$ production we find 
the diphoton rate $\sigma(pp\to\F\to\gamma\gamma) = (2.8\pm0.7)\fb$ at $s=(13\TeV)^2$, which is
  reproduced for 
\beq \frac{\Ggg}{M_\F}=\frac{s\sigma(pp\to\gamma\gamma)}{K_{gg}C_{gg}} = (3.8\pm 0.9)\,10^{-7},\eeq
where $K_{gg} C_{gg}=1.5\times 2140$ are partonic factors~\cite{1512.04933}.

In the limit of small $\beta_{\lambda_S}$
the diphoton mass and decay width are reproduced for
\beq
v_S=N \frac{310\GeV}{\sqrt{1+\tan^2\theta} }\sqrt{\frac{10^{-6}M_\F}{\Ggg}},\qquad
\beta_{\lambda_S} = \frac{5.9}{N^2} (1+\tan^2\theta)^2 \frac{\Ggg}{10^{-6} M}.\eeq
Even for $N=2$ this corresponds to a perturbative value of $g$, with $Ng^2/(4\pi)^2$
becoming smaller at larger $N$.

In figure \ref{run2} we provide a numerical example with $M_\F = 750\GeV$,
$M_{\F'} \simeq 1.7 \TeV$, $\Ggg \approx 3.5 \times 10^{-7} M_\F$, $M_1 \simeq M_{N-1} \approx 1.3 \TeV$. 
There are no Landau poles at energies much smaller than the Planck scale, and the stability conditions of eq.~(\ref{eq:V2p}) 
are violated only at low energy, when the desired Coleman-Weinberg mechanism takes place. 
We assumed central values for the SM parameters, and the Higgs quartic $\lambda_H$  remains positive up to the Planck scale, unlike in the SM:
the new dynamics eliminated the SM vacuum instability~\cite{1307.3536,SMinstab}.\footnote{Needless to say, new physics at the Planck scale
can give new sources of vacuum decay faster than the SM instability scale and faster than the diphoton lifetime.
The first part of this statement was strongly emphasised in~\cite{1307.5193}. 
For other diphoton models addressing the vacuum instability of the SM  see~\cite{vac750}.
}

\subsection{Model of type $\lambda$)}\label{FHH}
Models of type $\lambda$) --- those where the diphoton quartic is driven negative by RGE effects
of other quartics ---  need a multiplicity of $N$ scalars in order to avoid  Landau poles nearby by
sharing the needed relatively large quartics.
Some diphoton models introduce a multiplicity of $N$ states for a different reason: in order to 
mediate a sufficiently large $\F\to\gamma\gamma$ rate.
We thereby identify the charged particles that mediate $\F\to\gamma\gamma$ with the scalars that drive the diphoton quartic to negative values.

We now show that a successful model of this type is obtained by considering the massless limit
of the diphoton model proposed in~\cite{1602.01460}, where the multiplicity of $N$ states is justified by adding
an extra gauge group $\SU(N)$ with gauge coupling $g$.
The model employs the field content listed in table~\ref{tab:Mlambda}.
The three scalars are: the SM Higgs doublet $H$, a neutral singlet $S$ that will contain the diphoton $\F$, and
a charged and colored scalar $X$ in the fundamental $N$ of $\SU(N)$,
that mediates $\F\to\gamma\gamma, gg$ at one loop.

\begin{table}[t]
$$\begin{array}{|cc|cccc|}\hline
\hbox{name} &\hbox{spin}   & \SU(3)_c & \SU(2)_L & \U(1)_Y & \SU(N)  \\
\hline 
%H &0& 1& 2 & -1/2 & N   \\
X & 0 &3 &1 & Y &  N\\
S & 0 & 1 &1 & 0 &  1\\
\hline
{\cal N} & 1/2 & 1 &1 &0 & N\\
{\cal N}^c & 1/2 &1 &1 &0 &\bar N\\ \hline
\end{array}$$
\caption{\label{tab:Mlambda}\em Beyond the Standard Model
field content of the model of type $\lambda)$ of section~\ref{FHH}.}
\end{table}

Including the most generic dimension-less quartic couplings, the scalar potential of the model is:
\bea
 V(H, S,X) &=& \lambda_H |H|^4 - \lambda_{H S} |H|^2  S^2+
 \lambda_ S  S^4 + \lambda_{HX} |H|^2 |X|^2+ \nonumber \\
 && +\lambda_{X S}  S^2 |X|^2 +\lambda_X\Tr(XX^\dagger)^2+\lambda'_X \Tr(XX^\dagger XX^\dagger) 
 . \label{eq:V}\eea
The tree-level potential satisfies $V\ge 0$ when the quartic couplings satisfy \cite{Kannike:2012pe}
\beq\left\{
\begin{array}{l}
  \lambda_{S} > 0, \qquad  \lambda_{H} > 0, \qquad \lambda_{X} + \alpha \lambda'_{X} \ge 0,
  \\
  \bar{\lambda}_{HS} \equiv -\lambda_{HS} + 2 \sqrt{\lambda_{H} \lambda_{S}} \ge 0,
  \\
  \bar{\lambda}_{HX} \equiv \lambda_{HX} + 2 \sqrt{\lambda_{H} (\lambda_{X} + \alpha \lambda'_{X})} \ge 0,
  \\
  \bar{\lambda}_{XS} \equiv \lambda_{XS} + 2 \sqrt{\lambda_{S} (\lambda_{X} + \alpha \lambda'_{X})} \ge 0,
  \\
 \sqrt{\lambda_{H}} \lambda_{XS} + \sqrt{\lambda_{S}} \lambda_{HX} - \sqrt{\lambda_{X} + \alpha \lambda'_{X}} \lambda_{HS} 
  + 2 \sqrt{\lambda_{H} \lambda_{S}(\lambda_{X} + \alpha \lambda'_{X})} + \sqrt{\bar{\lambda}_{HS} \bar{\lambda}_{HX} \bar{\lambda}_{XS}} \ge 0,
\end{array}\right.\label{eq:V}\eeq
for $\alpha=1$ and for $\alpha=1/N_c$
(for $N\ge N_c=3$),
which are the extremal values of $$\alpha = \Tr(XX^\dagger XX^\dagger) /\Tr(XX^\dagger)^2.$$

A Coleman-Weinberg minimum is generated when the RGE running of the quartics
crosses the boundary of one of these conditions.
In practice, we are interested in the case where $\lambda_ S$ becomes negative while running to low energy,
while $\lambda_{H S}>0$ (such that the Higgs too acquires a vev) and $\lambda_{HX}>0$ (such that $X$ does not acquire any vev).

\medskip

Furthermore, the model contains $N_f$ extra fermions ${\cal N}\oplus {\cal N}^c$  with no SM gauge interactions and
 in the $N\oplus\bar N$ representation of $\SU(N)$.
 Such fermions receive mass
from  $ S {\cal NN}^c$ Yukawa couplings.
The lightest among them and among the $\SU(N)$ vectors are Dark Matter candidates.
The Yukawa couplings $X^* U{\cal N}$ (allowed if $Y=2/3$)
induces $X$ decays into SM up quarks and $N$.
Recasting LHC searches for similar objects~\cite{ATLAS}, we estimate a bound $M_X\circa{>}1.0 ~(1.2)\TeV$ on their masses for $N=3~(10)$.
We assume that such extra Yukawa couplings are small enough that we can neglect their contributions to the RGE.
In the presence of these extra fermions, the RGE
acquire infra-red fixed points which
allow the model to be RGE-extrapolated up to the Planck scale with a stable potential.

%where the range of the orbit space parameter $\alpha$ depends on the $SU(N)$ representation of $X$. The conditions are required to hold for both $\alpha = \alpha_{\rm min}$ (if $\lambda'_{X} > 0$) and $\alpha = \alpha_{\rm max}$ (if $\lambda'_{X} < 0$).
%\xxx{Is the range $\frac{1}{3} \leq \alpha \leq 1$ as implied by the below eq.?}
%The stability condition on $X$ is
%\beq \lambda_X+\lambda'_X>0,\qquad 3\lambda_X+\lambda'_X>0
%\eeq 

The RGE of the model are listed in appendix~\ref{RGElambda}.
They allow $\lambda_{H S}$ and $\lambda_{X S}$ to be naturally small, while $\lambda_{HX}\sim g_1^4$, $\lambda_ S\sim g^4$, $\lambda_H \sim g_2^4$.

%$X$ are assumed to be in the fundamental of an extra $\SU(N)$ gauge group, with gauge coupling $g$.

\subsubsection*{Masses}
%\beq (4\pi)^2 \beta_{\lambda_ S} =   N \lambda_{X S}^2+2\lambda_{H S}^2 + 72\lambda_{ S}^2.\eeq
The potential at one loop order can be approximated by inserting a running
$\lambda_ S$ in the tree-level potential of eq.\eq{V}:
\beq
\lambda_ S \simeq  \beta_{\lambda_S} \ln\frac{  S}{ S_*}  ,\eeq
where %$\beta_{\lambda_S} \approx \frac{1}{(4\pi)^2}\frac{9}{8} g_X^4$
$ S_*$ is the  scale below which $\lambda_ S$ becomes negative and
$\beta_{\lambda_ S} $ is given in eq.\eq{betalambdaF}. 
We can here neglect the running of the other couplings.
Expanding the scalars as
\beq  S = w + s,\qquad  H =\frac{1}{\sqrt{2}} \begin{pmatrix} 0 \cr v+h \end{pmatrix}\eeq
the effective potential is minimised by 
\beq
 w =  S_* \exp\bigg[-\frac{1}{4} +\frac{ \lambda_{H S}^2}{4\lambda_H \beta_{\lambda_ S}}\bigg],\qquad
v =  w \sqrt{\frac{\lambda_{HS}}{\lambda_H}}.\eeq
%\exp ( - \frac{1}{4} + \frac{\lambda_{HS}^2}{4\lambda_H \beta_{\lambda_S}}).\eeq
%The negligible second factor in the exponential recovers the full instability condition of eq.\eq{SB}.
The scalar mass matrix at the minimum in the ($h,s$) basis is
\beq 2v^2
\begin{pmatrix}
\lambda_H & - \sqrt{\lambda_H\lambda_{H S}} \cr
-\sqrt{\lambda_H\lambda_{H S}}  &  \lambda_{H S} + 2\beta_{\lambda_ S} {\lambda_H}/{\lambda_{H S}} 
\end{pmatrix}.
\eeq
The mass eigenstates are the physical Higgs and the diphoton $\F$.
In the limit of small  $\epsilon \equiv \lambda_{H S}^2/2\lambda_H\beta_{\lambda_ S}$, which corresponds to a Higgs mass smaller than the diphoton mass, 
the mass  eigenvalues  are
\beq M_h^2 \simeq   2 v^2 \lambda_H  ( 1-\epsilon+\cdots),\qquad
M_\F^2 \simeq  4 w^2  \beta_{\lambda_ S}( 1+\epsilon+\cdots).
%m_2^2 \simeq  v^2 ( \frac{2\beta_{\lambda_S}\lambda_H}{\lambda_{HS}}  +  \lambda_{HS} ),\qquad
\label{eq:eigen}
\eeq
%One has $M_h^2 \approx \lambda_{H S}  S^2$ and $M_X^2 = \lambda_{X S} S^2$
The diphoton mass
$M_\F$ is suppressed by a one loop factor because the diphoton is identified here with the pseudo-Goldstone boson of scale invariance.
The $h/\F$ mixing angle 
\beq \theta_{h\F} \simeq \frac{M_h^2}{M_\F^2}\frac{v}{w}\eeq
is below the experimental bound of eq.\eq{hFmixexp} for  $w\circa{>} 500\GeV$.
Finally,  $X$ acquires the mass
%The scalar masses in the leading order in the the small $\lambda_{H S}$ expansion are given by 
$ M_X \simeq  w \sqrt{\lambda_{X S}}$,
the extra fermions receive mass $M_{\cal N}=y_S w$ 
from  $y_S\, S  {\cal N} {\cal N}^c$ Yukawa couplings,
and the SM particles acquire the usual masses through the Higgs vacuum expectation value.

\subsubsection*{Diphoton rate}
The $\Ggg \equiv  \Gamma(\F\to\gamma\gamma)$ rate is given by
\beq \frac{\Ggg}{M_\F}=
\frac{9N^2 \alpha^2_{\rm em} Y^4 }{256\pi^3}\bigg|
 \frac{w \lambda_{X S} M_\F}{M_{X}^2}  F\left(\frac{4M_X^2}{M_\F^2}\right)
\bigg|^2 
\label{eq:gamrat}\eeq
where the loop function $F$ is
\beq F(x) = x\bigg[ x \arctan^2\left(\frac{1}{\sqrt{x-1}}\right) -1\bigg]  \stackrel{x\to\infty}{=} \frac{1}{3}  .\eeq
In the limit $M_X\gg M_\F/2$ the $\Ggg$ rate does not depend on $\lambda_{H S}$, as can be understood using 
 the Low Energy Theorem of eq.\eq{LET}, taking into account that, in the present model,
 $\wp=X$ with $\sfrac{M_X(\F)}{M_X} =\F/w + \cdots$.

\begin{figure}
$$\includegraphics[width=0.44\textwidth]{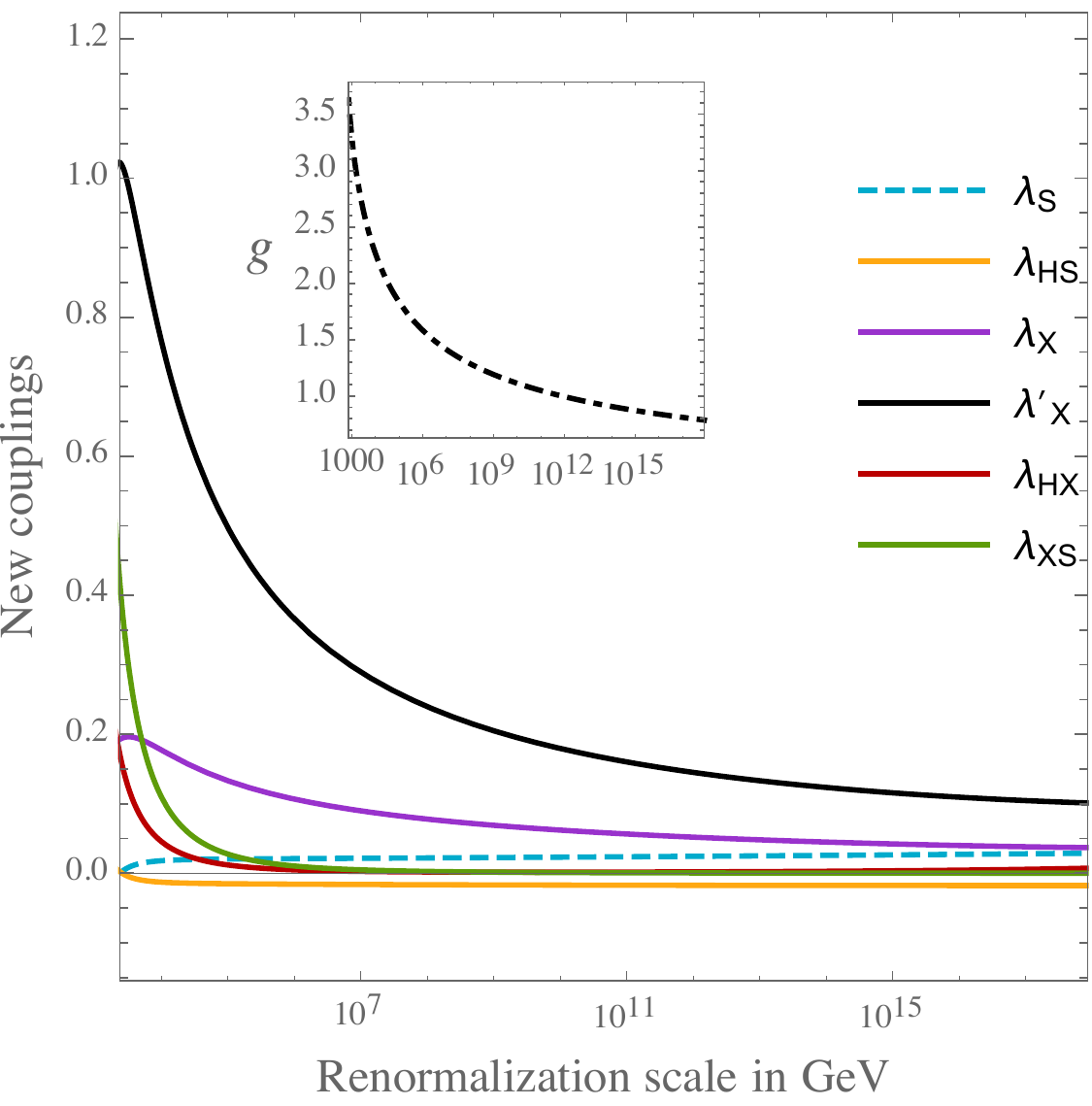}\qquad\qquad
\includegraphics[width=0.44\textwidth]{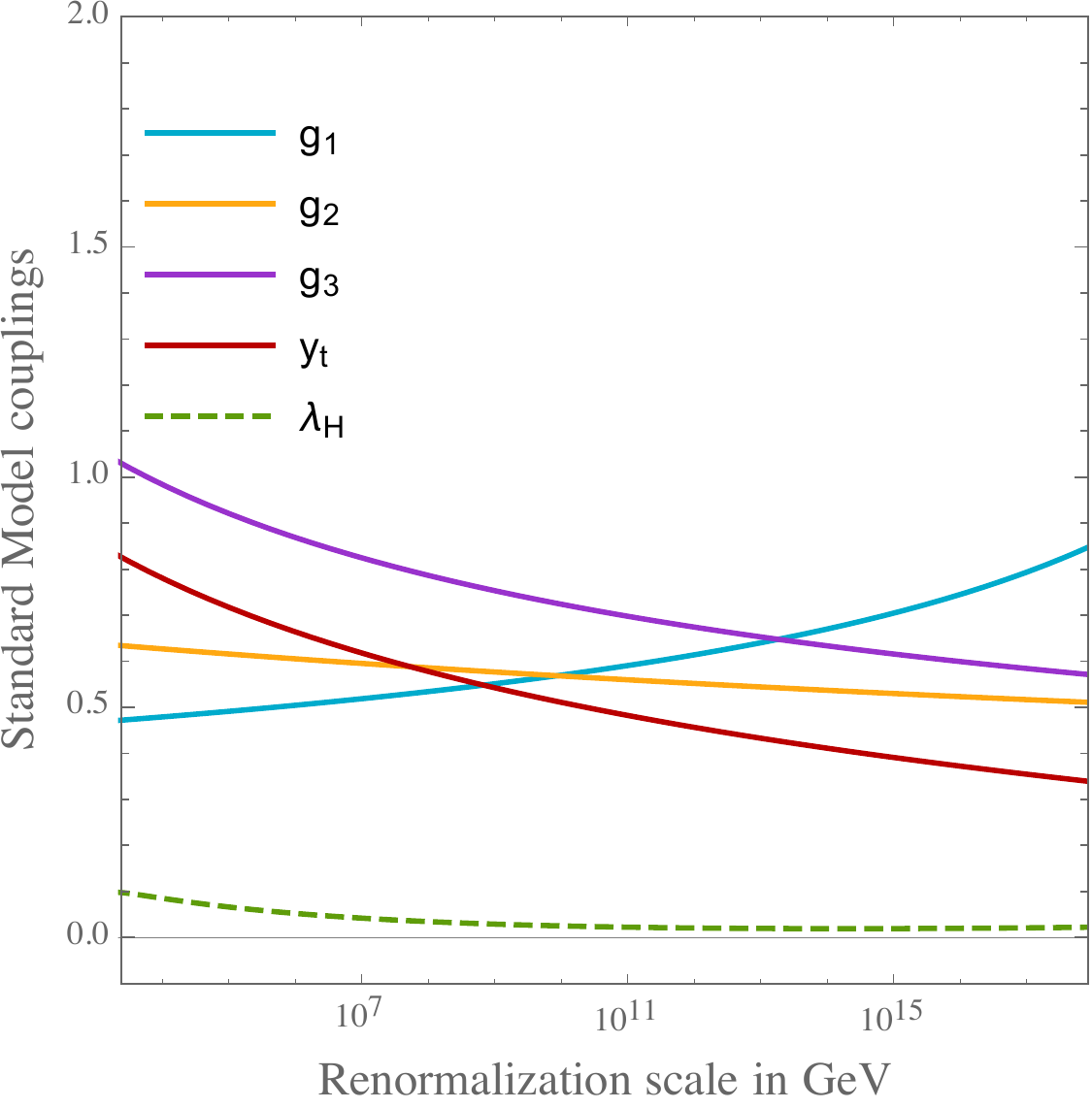}$$
$$\includegraphics[width=0.44\textwidth]{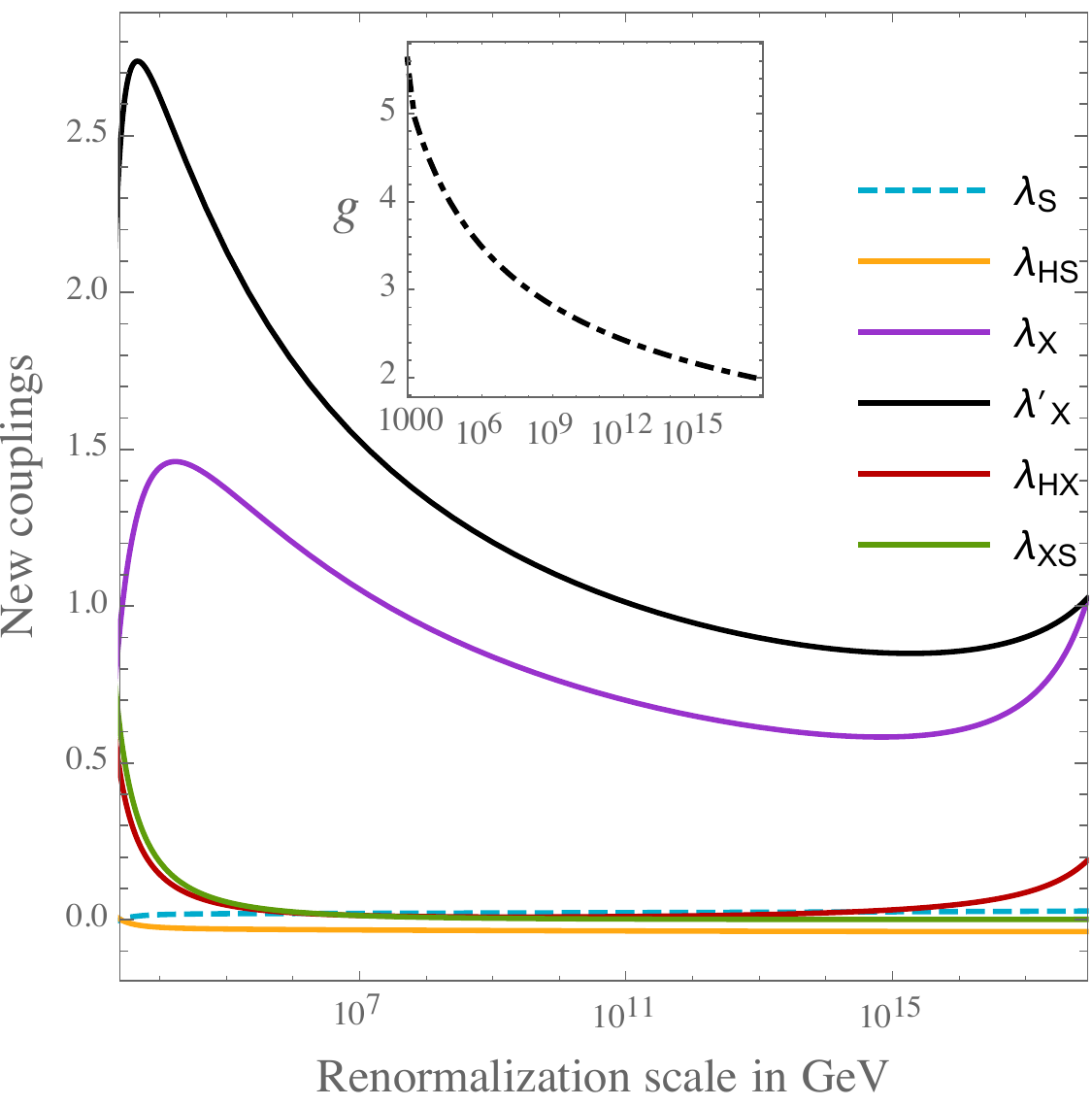}\qquad\qquad
\includegraphics[width=0.44\textwidth]{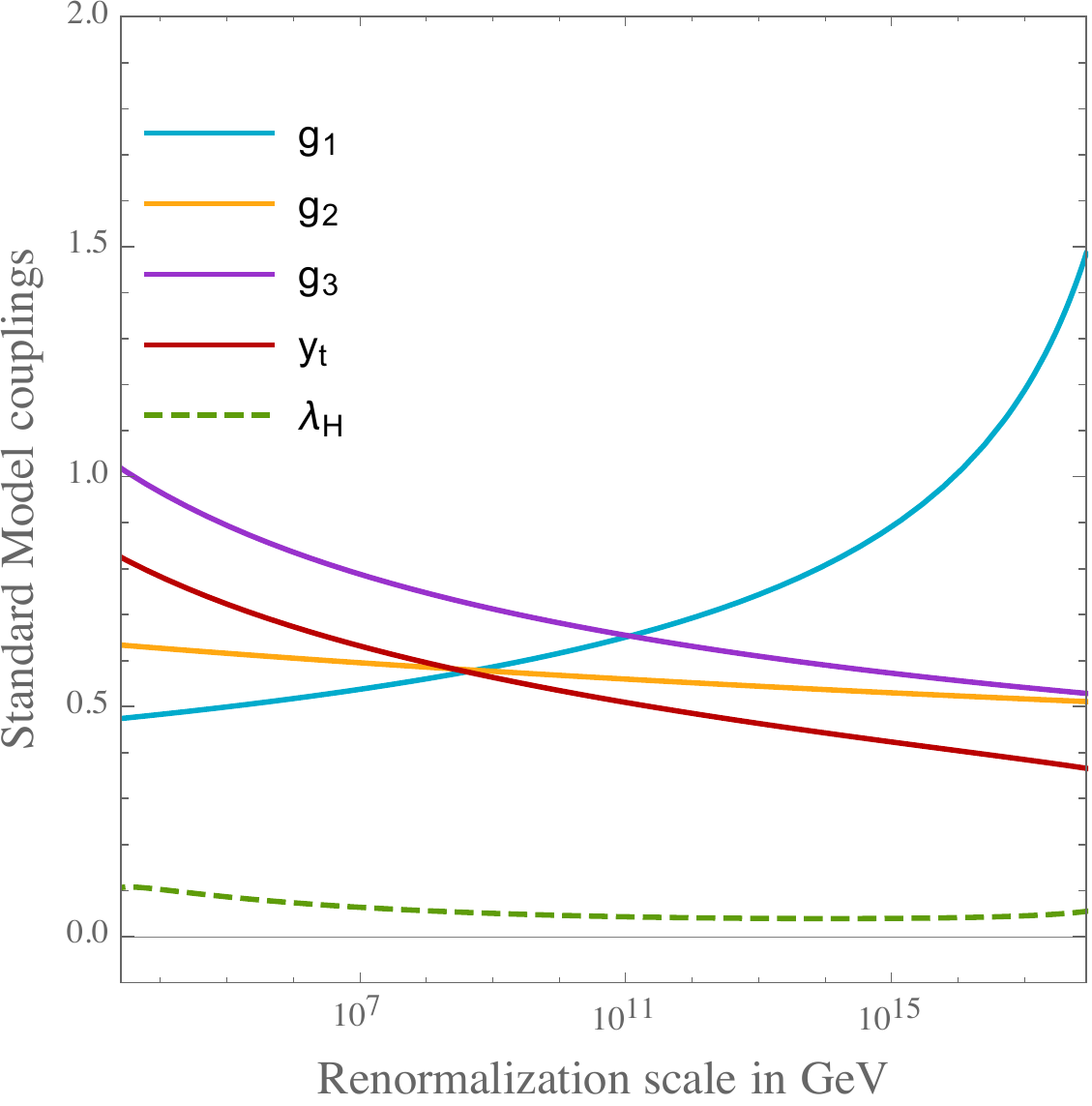}$$
\caption{\em 
\label{fig:Yuk} Running of the couplings for  central values for the SM parameters, that we took from \cite{1307.3536}, and $g^2(M_\F) = (4\pi)^2/N$. We find that  that varying the values of $\lambda_X(M_\F)$, $\lambda_X'(M_\F)$,  $\lambda_{HX}(M_\F)$ have a very small impact on the running at energies much bigger than  $\F_*$. {\bf Upper plots:} We assumed  $Y=2/3$, $N=12$, $w=1890\GeV$, $5N$ fermions in the  $N\oplus\bar N$ representation of $\SU(N)$; we also take $\lambda_X(M_\F)=0.15$, $\lambda_X'(M_\F)=0.28$,  $\lambda_{HX}(M_\F)=0.17$. {\bf Lower plots:} We assumed  $Y=4/3$, $N=5$, $w=2000\GeV$, 26 fermions in the  $N\oplus\bar N$ representation of $\SU(N)$; we also take $\lambda_X(M_\F)=0.15$, $\lambda_X'(M_\F)=0.28$,  $\lambda_{HX}(M_\F)=0.47$.
\label{run}}
\end{figure}

\medskip
 
 As usual a large multiplicity $N$ enhances $\Ggg$; in our context it also enhances $M_\F$.
In order to reproduce desired values of 
$M_h=125\GeV$, $M_\F=750\GeV$, $M_X$ and $\Ggg$,
the model parameters are fixed to the following values 
\beq\begin{array}{ll}\displaystyle
\lambda_{X S} \approx  0.24 Y^4 \bigg(\frac{\TeV}{M_X}\bigg)^6 \bigg(\frac{10^{-6}}{\Ggg/M_\F}\bigg),\qquad & \displaystyle
 w \approx \frac{2.0\TeV}{Y^{2}}\bigg(\frac{M_X}{\TeV}\bigg)^4 \bigg(\frac{\Ggg/M_\F}{10^{-6}}\bigg)^{1/2},\\[4mm]  \displaystyle
 \lambda_{H S}\approx0.002\, Y^4\bigg(\frac{10^{-6}}{\Ggg/M_\F}\bigg),& \displaystyle
N \approx \frac{30}{Y^{4}}\bigg(\frac{M_X}{\TeV}\bigg)^4 \bigg(\frac{\Ggg/M_\F}{10^{-6}}\bigg).
\end{array}
\eeq
at leading order in $M_h/M_\F\ll 1$, and neglecting higher order corrections such as the running of the coupling constants between $M_\F$ and $S_*$.
We see that $N$  is large (tens) for $Y=2/3$
and depends strongly on $Y$, such that a small $N$ is obtained for $Y=4/3$ or $5/3$.

\subsubsection*{Numerical examples}
In figure \ref{run} we provide two numerical examples with $M_\F = 750\GeV$,
 with no Landau poles at energies much smaller than the Planck scale and with all the stability conditions of eq.~(\ref{eq:V}) satisfied. 
We assumed central values for the SM parameters, and the Higgs quartic $\lambda_H$  remains positive up to the Planck scale unlike in the SM.

In the upper example we have $Y=2/3$ and $M_X\approx  1.2\TeV$
The example leads to $\Ggg/M_\F \simeq 3.1\times 10^{-7}$ and to a
 diphoton decay rate in two gluons
$\Gamma_{gg}/M_\F \simeq 4.7 \times 10^{-5} $.
%A nice feature of t
This example  employs $N=12$, which is compatible 
%with $\F$-theory~\cite{SU27}and 
with experimental bounds.

Given that such a large $N$ might look implausible from a  low-energy perspective, we provide in the lower row a second example
with $N=5$, achieved by increasing $Y=4/3$.
This example has  $M_X= 1.6 \TeV$, $\Ggg \simeq 4.4\times 10^{-7} M_\F$, $\Gamma_{gg} \simeq 4.2 \times 10^{-6} M_\F$.

In both cases  the $\SU(N)$ gauge constant becomes relatively large at the diphoton scale such that
the couplings (in particular $\lambda_{XS}$) have a fast running; we tried to include such corrections
by renormalising couplings at appropriate scales.

\section{Strongly coupled models}\label{strong}
In this section we try to build dimensionless models where a new gauge interaction
(we use for it the old-fashioned name TechniColor, or TC) 
becomes strong around the weak scale, inducing the weak scale  and the diphoton $\F$.
Like in the weakly-coupled case, it is easy to reproduce the diphoton excess by adding one extra ad-hoc  diphoton scalar 
(see~\cite{1604.07712}  for one such model).
We are interested in models where the diphoton automatically emerges as a bound state of the TC dynamics.
In view of strong LHC bounds on extra bound states, especially the ones with color,
plausible models identify the diphoton with a bound state that is much lighter than the others.
The diphoton could be a TC$\eta$ (here discussed in section~\ref{TCeta}) or a
TC-dilaton (here discussed in section~\ref{TCd}).
The TC-dilaton is especially interesting from our point of view, given that it is the pseudo-Goldstone boson of scale invariance.

\begin{table}
$$\begin{array}{|cc|cccc|}\hline
\hbox{name} &\hbox{spin}   & \SU(3)_c & \SU(2)_L & \U(1)_Y & \SU(N)  \\
\hline 
{\cal Q} & 1/2 & 3 &2 &+1/6 & N\\
{\cal Q}^c & 1/2 &  \bar 3 &2 &-1/6 &\bar N\\
{\cal U} & 1/2 & \bar 3 &1 &-2/3 & N\\
{\cal U}^c & 1/2 &  3 &1 &+2/3 &\bar N\\ \hline
\end{array}$$
\caption{\label{tab:M3}\em Beyond the Standard Model
field content of the strongly-coupled model of section~\ref{TCeta}.}
\end{table}

\subsection{The diphoton as a TC$\eta$}\label{TCeta}
A class of strongly-coupled models that aims at reproducing the diphoton excess
are those where the diphoton $\F$ is identified with a TC$\eta$ composite pseudo-scalar,
given that this field is a light pseudo-Goldstone boson with anomalous couplings to SM vectors.
The various models proposed in the literature~\cite{1512.04933,Fstrong} employ massive techni-quarks, 
and a massive SM Higgs doublet.
The masses are assumed to be comparable to the TC scale, but the coincidence is left unexplained.

Our goal is exploring  whether such masses can be all set to zero,  obtaining a dimension-less model.
For concreteness, we consider a model with $G_{\rm TC}=\SU(N)$ and the techni-quark
content of table~\ref{tab:M3}\footnote{A similar model with up-type quarks replaced by down-type quarks 
has been considered in~\cite{1503.08749} because the lightest TCbaryon is a good Dark Matter candidate.}
%\beq\Q = Q\oplus Q^c\oplus U\oplus U^c
%%\qquad\hbox{or}\qquad
%%\Q = Q\oplus Q^c\oplus D\oplus D^c.
%\eeq
which allows for two Yukawa couplings to the SM Higgs doublet $H$:
\beq y_1 H {\cal Q U} + y_2 H^\dagger  \Q^c  {\cal U}^c  + \hbox{h.c.}= H \bar\Psi_\Q (y + i\gamma_5 \tilde y)\Psi_{\cal U} + \hbox{h.c.}\eeq
In the latter expression we introduced Dirac spinors $\Psi$ and 
the scalar coupling $y = (y_1+y_2^*)/2$ and the pseudo-scalar coupling 
$\tilde y =i (y_1- y_2^*)/2$,
such that $|y|\ll |\tilde y|$ and $|\tilde y| \ll |y|$ are radiatively stable special cases, that 
can be justified by assuming a CP-like symmetry.

\medskip

We now discuss the composite states.
Among the many states around the TC$\rho$ mass,
$m_\rho \sim g_\rho f_{\rm TC}$ with $g_\rho \sim 4\pi/\sqrt{N}$,
there is the TC $\eta' \sim QQ^c+UU^c$ singlet which receives a mass from TC anomalies.
The TC dynamics breaks the global accidental symmetry
$\SU(9)_L\otimes\SU(9)_R\to\SU(9)_V$ giving 80 lighter techni-pions, with the following
SM quantum numbers:
\beq (1,1)_0 \oplus (1,2)_{\pm 1/2}  \oplus (1,3)_0 \oplus 2(8,1)_0 \oplus (8,2)_{\pm 1/2}  \oplus (8,3)_0.\eeq
Taking into account that we assume massless techni-quarks, the techni-pions consist of:
\begin{itemize}
\item[72)] 9 color octets, $Q Q^c$, $UU^c$, $QU^c$, $Q^cU$, which get positive squared masses at loop level
from QCD interactions,
$m\approx \sqrt{\frac{3}{4\pi} \alpha_3 C}  m_\rho^2\sim 0.2m_\rho$, where
$C=(N^2-1)/2N$;

\item[3)] a $\SU(2)_L$ triplet $QQ^c$, which similarly gets a positive squared mass from weak interactions;

\item[1)] a TC $\eta\sim UU^c-\frac12 QQ^c$ singlet (to be identified with the diphoton),
which gets a squared mass from Yukawa loop corrections with unknown sign and
of order
$m_\eta \sim y m_\rho/4\pi$, see appendix~A.3 of~\cite{1410.1817};

\item[4)] a TCpion $\pi_2$  from $QU^c, Q^c U$ 
with the same gauge quantum numbers as the Higgs doublet forms whenever gauge quantum numbers allow
for a Yukawa coupling to $H$.
It gets a mass 
$m_{\pi_2} \sim g_2 m_\rho/4\pi$ from weak gauge interactions,
plus another contribution from $y_{1,2}$.

\end{itemize}
Furthermore, $\pi_2$ acquires a tree level mixing with the Higgs boson.  The resulting mass matrix is
\beq\bordermatrix{ & \pi_2^* & H^* \cr 
\pi_2 &  ({\cal O}(g_2^2) \pm {\cal O}(y^2))/(4\pi)^2 & {\cal O}(y) \sqrt{N}/(4\pi) \cr 
H &  {\cal O}(y) \sqrt{N}/(4\pi) &-{\cal O}( y^2)  N/(4\pi)^2}  \label{pihmatrix}  m_\rho^2
 \eeq
 The tree-level mixing induces a negative see-saw contribution
$\Delta m^2_H \sim -\tilde y^2 m_\rho^2 f_{\rm TC}^2/m_{\pi_2}^2$ to the Higgs mass parameter,
where .
% (see appendix A.2 of~\cite{1410.1817}).
This contribution is too big, given that the diphoton only gets a mass at loop level.

\begin{figure}[t]
$$\includegraphics[width=0.93\textwidth]{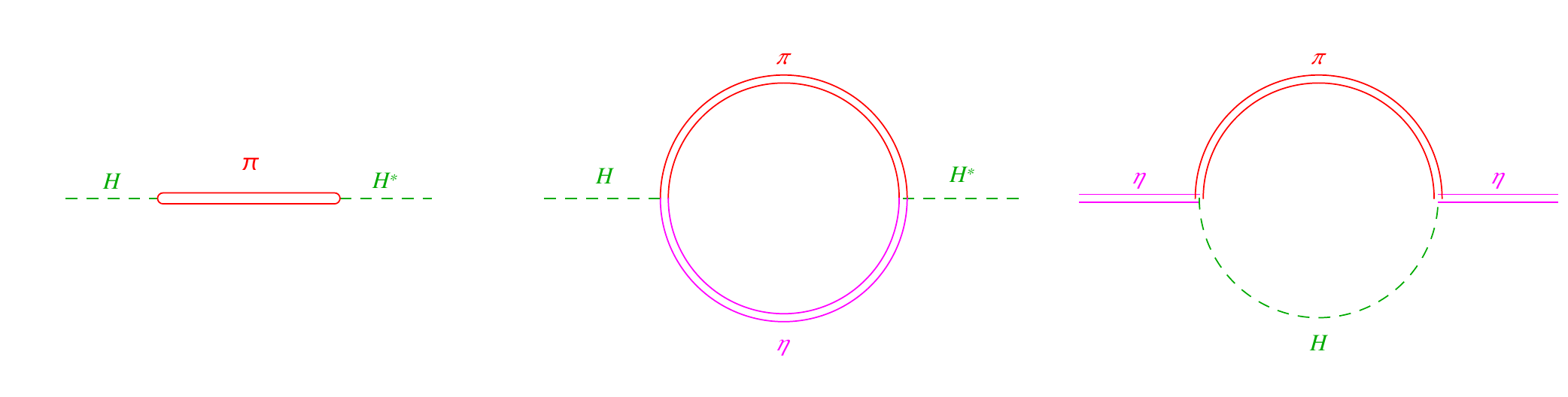}$$
\caption{\em 
\label{fig:YukTC} Correction to the Higgs and diphoton mass coming from the Yukawa couplings.}
\end{figure}

Such tree-level contribution vanishes if we assume $\tilde y =0$ -- a
natural special case that respects a CP-like parity.
Indeed, $H$ is a scalar, while $\pi_2$ and $\eta$ are pseudo-scalars.
Their tree-level potential includes a CP-conserving term 
$ \sim y m_\rho  H \pi_2^* \eta + \hbox{h.c}$.
As a result, both $H$ and of $\eta$ receive loop-level masses of order $y m_\rho/4\pi$,
with no symmetry relation among them, given that $H$ is elementary while $\eta$ is composite,
see fig.\fig{YukTC}.
With a relatively large value of $y\sim \hbox{few}$ the model can give 
$M_\F\approx 750\GeV$ together with $f_{\rm TC}\approx  100\GeV\times N$ as demanded by the diphoton rate suggested by preliminary ATLAS and CMS results.
The $\eta$ eigenstate has a vanishing color anomaly (see table~2 of~\cite{1602.07297});
this is not a problem because, taking into loop corrections, 
the mass eigenstate $\F\approx \eta + \eta' y^2/(4\pi)^2$ acquires a sufficiently large coupling to gluons.

\subsection{The diphoton as a TC-dilaton or TC$\sigma$}\label{TCd}
In strongly coupled TC models with a QCD-like dynamics
the mass of the pseudo-Goldstone boson of scale invariance
is not suppressed with respect to the other composite states, because strong interactions give a fast running that strongly breaks scale invariance.
On the other hand, such state is somehow lighter than the other bound states
in models with a `walking' dynamics
(namely, a $\beta$ function of the new gauge interaction which remains somewhat small), 
which is obtained in TC models with a larger matter content than QCD.
This lightness is beneficial for the diphoton phenomenology, given that other colored bound states 
must be heavier  than $750\GeV$ in order to satisfy LHC bounds.

\medskip

From a low-energy  perspective, such light state is the $\sigma$ field sometimes explicitly included in effective chiral Lagrangians,
where its vacuum expectation value,  $\sigma = f_{\rm TC}+\F$,
breaks scale invariance as well as a global chiral symmetry,
such that $f_{\rm TC}$  becomes the techni-pion decay constant.
See~\cite{1603.05668} for a recent discussion of the $\sigma$, and~\cite{1512.04933,dilaton} for recent discussions of the dilaton.
In the limit where $\F$ is lighter than the other bound states, its coupling to SM vectors
is dictated by eq.\eq{LET}: using $\ln\sfrac{M_\wp(\F)}{M_\wp}\simeq \F/f_{\rm TC}$
 we find
\beq  \label{eq:techniGgg}
\frac{\Ggg}{M_\F} = 10^{-6} \bigg(\frac{120\GeV}{f_{\rm TC}}\bigg)^2 \Delta b_{\rm em}^2\eeq
where $ \Delta b_{\rm em}$ is the techniquark contribution to the running of the electromagnetic coupling.
For example, a color triplet techni-fermion with hypercharge $Y$ 
(such as ${\cal U}\oplus {\cal U}^c$ in table~\ref{tab:M1})
contributes  as $ \Delta b_{\rm em}=4N Y^2$, while a techni-scalar contributes 4 times less.
The SM $\Delta b_{Y,2,3}$ must be smaller than about 10, in order to avoid sub-Planckian Landau poles
for the SM gauge couplings.

\medskip

The dynamically generated Higgs mass depends on the model.
In the minimal case where the Higgs has no direct coupling to techni-particles,
electro-weak loop effects induce a contribution to its squared mass of order
$M_h^2 \sim - \alpha_Y^2 f^2_{\rm TC}$~\cite{1410.1817},
which is negative but small, in view of eq.\eq{techniGgg}.
A larger model-dependent tree-level contribution is obtained if  the Higgs has a Yukawa coupling to techni-fermions or a
quartic coupling to techni-scalars.

\section{Conclusions}\label{concl}
%One can also search for models (weak or strong) where 
%the extra fermions are chiral under an extra U(1) or SU(2);
%$\F$ can decay into the vectors.

%We therefore study models where one of the required new particles can be identified with a new resonance $\digamma$ decaying into two   photons. 

 We proposed dimension-less models where a new resonance $\digamma$ decaying into two   photons 
dynamically breaks scale invariance, generating a weak scale hierarchically smaller than the Planck scale.
The diphoton channel is interesting as it is particularly clean and has therefore a great discovery potential.
All particles acquire their masses from their couplings to $\F$: thereby
the smoking gun of this scenario is observing that $\F$ couples to all particles proportionally to their mass.

As a benchmark case we identified $\F$ with  the 750 GeV resonance hinted by LHC data.  Although it could very well be that this excess is a statistical fluctuation, it nevertheless provides an interesting example. 
The $\F\to\gamma\gamma$ rate suggested by LHC data
is obtained adding extra charged particles, 
heavier than the SM particles and thereby more strongly coupled to $\F$.

Diphoton models of this type generically need that such extra particles have masses around the weak scale.
Unlike the SM fermions, such extra particles have no chirality reason to be around the weak scale.
Scale invariance provides one possible reason: like the Higgs, these extra charged particles acquire a mass from $\F$.
All particles are massless until the diphoton develops a vev or condensate $w$.

A generic scalar that acts as the `Higgs of the Higgs' can have a $w$ and/or a mass $M$ much larger than the weak scale,
provided that it is very weakly coupled to the Standard Model (SM).
This is not possible if such scalar is identified with the diphoton: $w$ and $M$ are now
fixed by the diphoton mass and rate in $\gamma\gamma$.

\medskip

In section~\ref{FN} we presented a weakly-coupled dimension-less model where the diphoton is charged under extra gauge interactions,
which induce the Coleman-Weinberg mechanism.
In section~\ref{FHH} we presented a weakly-coupled dimension-less model where the Coleman-Weinberg mechanism is induced by
quartic interactions of the diphoton with the charged scalars that mediate $\F\to\gamma\gamma$.

Both models can be RGE-extrapolated up to the Planck scale, thereby generating the large hierarchy with respect to the weak scale.
Both models contain Dark Matter candidates.
Both models remove the instability of the SM potential.
Both models give rise to an extended phase transition when the diphoton and the Higgs acquire vacuum expectation values.
Such phase transition can be of first order,
possibly giving gravitational wave signals~\cite{1604.05035} and the baryon asymmetry~\cite{1407.0030}.

\medskip

In section~\ref{TCeta} we discussed dimension-less strongly-coupled models where the diphoton is a pseudo-scalar bound state analogous of the $\eta$ in QCD.
In section~\ref{TCd} we discussed dimension-less strongly-coupled models where the diphoton is a scalar bound state analogous of the $\sigma$ in chiral effective Lagrangians,
or of a dirty dilaton in fundamental strongly coupled models with walking dynamics.
Both scenarios can produce a weak scale lighter than the diphoton mass by $M_h/M_\F\sim1/6$, but this needs extra model building features.

\footnotesize

\subsection*{Acknowledgements}
This work was supported by the grant 669668 -- NEO-NAT -- ERC-AdG-2014.
Kristjan Kannike was supported by the Estonian Research Council grant PUT799, the grant IUT23-6 of the Estonian Ministry of Education and Research, and by the EU through the ERDF CoE program.
We thank Roberto Franceschini and Francesco Riva for useful discussions.

\small

\appendix

\section{RGE for model $\mb{g}$)}\label{RGEg}

Defining $\beta_{g} = dg/d\ln\mu$,
the one-loop RGE for the new couplings are
\begin{eqnsystem}{sys:RGE}
(4\pi)^2 \beta_{\lambda_S} &=&   4(4+N)\lambda_{S}^2+\frac{3(N-1)(N^2+2N-2)}{4N^2} g^4+ \nonumber\\
&&+2 (\lambda_{SX}^2+\lambda_{HS}^2)-6 y_S^4+\lambda_S\bigg[12 y_S^2 - 6 \frac{N^2-1}{N}g^2\bigg]
, \label{eq:betalambdaF2}\nonumber\\
(4\pi^2)\beta_{\lambda_{HS}}&=&12 y_S^2 y_t^2-4 \lambda_{HS}^2-4 \lambda_{HX} \lambda_{SX}+\\
&&+\lambda_{HS} \left[-3 g^2\frac{N^2-1}{N}-\frac{9 g_1^2}{10}-\frac{9 g_2^2}{2}+6 y_S^2+6 y_t^2+12 \lambda_{H}+(4 N+4) \lambda_{S}\right], \nonumber\\
(4\pi^2)\beta_{\lambda_{HX}}&=&-8 \lambda_{HX}^2-2 N \lambda_{HS} \lambda_{SX}+\\
&&+\lambda_{HX}\left[-\frac{9 g_1^2}{10}-\frac{9 g_2^2}{2}+6 y_t^2+12 N y_X^2+12 \lambda_{H}+24 \lambda_{X}\right], \\
(4\pi^2)\beta_{\lambda_{SX}}&=&12 y_S^2 y_X^2-8 \lambda_{SX}^2-4 \lambda_{HS} \lambda_{HX}+\nonumber\\
&&+\lambda_{SX} \left[-3 g^2\frac{N^2-1}{N}+6 y_S^2+12 N y_X^2+(4 N+4) \lambda_{S}+24 \lambda_{X}\right],\\
(4\pi^2)\beta_{\lambda_{X}}&=&-6 N y_X^4+24 N \lambda_{X} y_X^2+2 \lambda_{HX}^2+N \lambda_{SX}^2+72 \lambda_{X}^2,\\
(4\pi^2)\beta_{y_X}&=&(6 N+3) y_X^3+ y_X\left[-3 \frac{N^2-1}{N} g^2-\frac{8 g_1^2}{5}-8 g_3^2+\frac{y_S^2}{2}\right], \\
(4\pi^2)\beta_{y_S}&=&\frac{{N+7}}{2} y_S^3+ y_S\left[-\frac{3}{2} g^2\frac{N^2-1}{N}-\frac{8 g_1^2}{5}-8 g_3^2+y_t^2+\frac{y_X^2}{2}\right],\\
(4\pi^2)\beta_{g}&=&\left(\frac{13}{6}-\frac{11 N}{3}\right) {g}^3.\\[2mm]
\riga{Finally, the RGE for the SM couplings are}\\
(4\pi^2)\beta_{g_1}&=&\left(\frac{41}{10}+\Delta b_1\right) g_1^3,\qquad \Delta b_1 = \frac{16 N}{15}, \label{eq:db1}\\
(4\pi^2)\beta_{g_2}&=&-\frac{19 g_2^3}{6},\\
(4\pi^2)\beta_{g_3}&=&\left(-7+\Delta b_3\right) g_3^3,\qquad \Delta b_3=\frac{2 N}{3}, \label{eq:db3}\\
(4\pi^2)\beta_{y_t}&=&\frac{9 y_t^3}{2}+ y_t\left[-\frac{17 g_1^2}{20}-\frac{9 g_2^2}{4}-8 g_3^2+\frac{N y_S^2}{2}\right], \\
(4\pi^2)\beta_{\lambda_{H}}&=&\frac{27 g_1^4}{200}+\frac{9 g_2^2 g_1^2}{20}+\frac{9 g_2^4}{8}-6 y_t^4+24 \lambda_{H}^2+N \lambda_{HS}^2+2 \lambda_{HX}^2+\\
&&+ \lambda_{H}\left[12 y_t^2-\frac{9 g_1^2}{5}-9 g_2^2\right] .
\end{eqnsystem}

\section{RGE for model $\mb{\lambda}$)}\label{RGElambda}
Defining $\beta_{g} = dg/d\ln\mu$,
the RGE for the new couplings are
\begin{eqnsystem}{sys:RGE}
(4\pi)^2 \beta_{\lambda_ S} &=&   72\lambda_{ S}^2+3N \lambda_{X S}^2 +2\lambda_{H S}^2, \label{eq:betalambdaF}\\ 
(4\pi)^2 \beta_{\lambda_X} &=&   4(3 N+4) \lambda_X^2 +12\lambda_X'^2+ 2\lambda_{X S}^2+ 2\lambda_{HX}^2+ \nonumber\\
&&
+\lambda_X\bigg[ 8(3+N)\lambda_X' - \frac{6(N^2-1)}{N} g^2   -16 g_3^2-\frac{36Y^2 g_1^2}{5} +\frac{54 Y^4 }{25}g_1^4\bigg]+\\
&&+   \frac{3 (N^2 + 2)}{4N^2} g^4 +\frac{11}{12}g_3^4 +\frac{3 N + 1}{N}g^2g_3^2 
-\frac{18Y^2g^2g_1^2}{5N}-\frac{6Y^2 g_3^2g_1^2}{5} ,\nonumber \\
(4\pi)^2 \beta_{\lambda_X'} &=&
4(3+N)\lambda_X'^2+\lambda_X' \bigg[24\lambda_X-\frac{36Y^2 g_1^2}{5} - \frac{6(N^2-1)}{N} g^2 -16 g_3^2\bigg]+\nonumber \\ 
&&
+\frac{3 (N^2 - 4)}{4 N}g^4+\frac{5g_3^4}{4}-\frac{N + 3}{N}g^2g_3^2+   \frac{18 Y^2g^2g_1^2}{5}+\frac{18Y^2 g_3^2g_1^2}{5}
 ,\\
(4\pi)^2 \beta_{\lambda_{X S}} &=&  
8\lambda_{X S}^2-4\lambda_{H S}\lambda_{HX} +4 \lambda_{X S} \bigg[(1+3N)\lambda_X+(3+N)\lambda'_X+\nonumber\\
&&
+6\lambda_ S-\frac{9Y^2 g_1^2}{10}- \frac{3(N^2-1)}{4N}  g^2-2 g_3^2 \bigg],\label{eq:RGEXF}\\
(4\pi)^2 \beta_{\lambda_{HX}} &=& \lambda_{HX} \left[4(1+3N) \lambda_X+4(3+N) \lambda'_X-\frac{(36 Y^2+9)g_1^2 }{10}-\frac{9 g_2^2}{2}+12 \lambda_H+6 y_t^2\right]+\nonumber\\ 
&& - 4 \lambda_{H S} \lambda_{X S}+4 \lambda_{HX}^2 + \frac{27 g_1^4 Y^2}{25} - \frac{3(N^2-1)}{N} \lambda_{HX} g^2- 8 \lambda_{HX} g_3^2,\\
(4\pi)^2 \beta_{\lambda_{H S}} &=&  \lambda_{H S}\bigg[24\lambda_ S-\frac{9g_1^2}{10}-\frac{9g_2^2}{2}+6y_t^2+12 \lambda_H\bigg]-6N \lambda_{X S} \lambda_{HX} -8 \lambda_{H S}^2,\\
(4\pi)^2 \beta_{g}&=&  g^3 \left( -\frac{11}{3} N +\frac12+\frac23 N_f \right ).\\[2mm]
\riga{Finally, the RGE for the SM couplings are}\\
(4\pi)^2 \beta_{g_1} &=& g_1^3\frac{41+6NY^2}{10}, \\
(4\pi)^2 \beta_{g_2}&=&-\frac{19 g_2^3}{6},\\
(4\pi)^2 \beta_{g_3}&=& g_3^3\left(- 7 + \frac{N}{6}\right),\\
(4\pi)^2 \beta_{\lambda_{H}} &=&2 \lambda_{H S}^2+3N \lambda_{HX}^2+ \frac{27 g_1^4}{200}+\frac{9 g_1^2 g_2^2}{20}+\frac{9 g_2^4}{8}+\nonumber \\ &&+\lambda_{H} \left(-\frac{9 g_{1}^2}{5}-9 g_2^2+12 y_t^2\right)+24 \lambda_H^2-6 y_t^4,\\
(4\pi)^2 \beta_{y_t} &=&y_t\left(\frac92 y_t^2-\frac{17g_1^2}{20} -8g_3^2-\frac{9g_2^2}{4}\right).
\end{eqnsystem}

\end{document}